\documentclass[twocolumn,prl]{revtex4}

\usepackage{graphicx}
\usepackage{amsmath,amssymb}
\usepackage{textcomp}

\newcommand{\bra}[1]{\langle #1 |} 
\newcommand{\ket}[1]{\,| #1 \rangle} 

\mathchardef\mhyphen="2D

\begin{document}

\bibliographystyle{naturemag}

\title{Quantum Process Tomography of an Optically-Controlled Kerr Non-linearity}

\author{Connor Kupchak, Samuel Rind, Bertus Jordaan and Eden Figueroa}
\affiliation{Department of Physics and Astronomy, Stony Brook University, New York 11794-3800, USA}

\begin{abstract}
Any optical quantum information processing machine would be comprised of fully-characterized constituent devices for both single state manipulations and tasks involving the interaction between multiple quantum optical states. Ideally for the latter, would be an apparatus capable of deterministic optical phase shifts that operate on input quantum states with the action mediated solely by auxiliary signal fields. Here we present the complete experimental characterization of a system designed for optically controlled phase shifts acting on single-photon level probe coherent states. Our setup is based on a warm vapor of rubidium atoms under the conditions of electromagnetically induced transparency with its dispersion properties modified through the use of an optically triggered N-type Kerr non-linearity. We fully characterize the performance of our device by sending in a set of input probe states and measuring the corresponding output via balanced homodyne tomography and subsequently performing the technique of coherent state quantum process tomography. This method provides us with the precise knowledge of how our optical phase shift will modify any arbitrary input quantum state engineered in the mode of the reconstruction.
\end{abstract}

\maketitle

Emergent quantum optical technologies that will one day comprise the heart of future quantum networks and computers will necessitate that their performance be understood precisely. Furthermore, these devices will require the capacity for both single and multi-mode quantum processing. While a majority of recent experimental advancements regarding light-matter interaction at the quantum level have focused on operations involving a single quantum optical state \cite{Lvovsky2009,Bussieres2013,Northup2014,Reiserer2014}, further functionalities are necessary in order to implement a future quantum processing machine. Namely the creation of fully-characterized quantum light-matter interfaces suited for the interaction between weak quantum optical states and triggering signal fields.

\begin{figure*}[]
\centerline{\includegraphics[width=2.0\columnwidth]{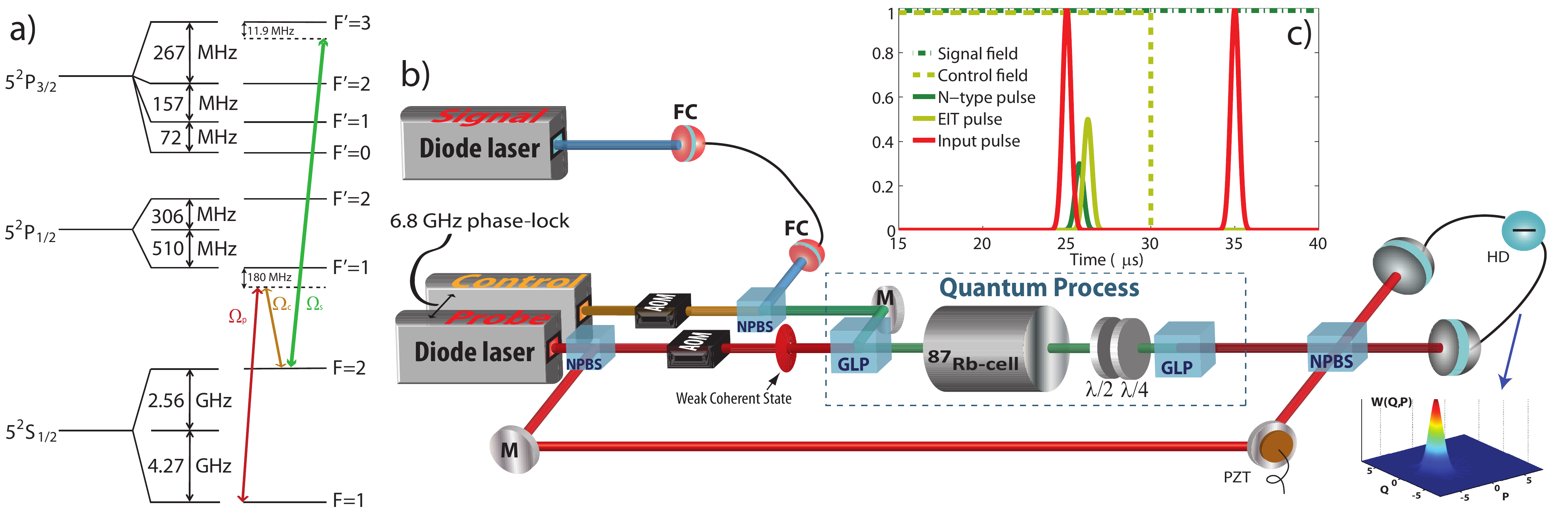}}
\caption{\textbf{Atomic level scheme and experimental setup.} (a) Atomic level diagram for the N-type scheme in $^{87}$Rb. (b) Experimental setup for quantum process tomography of a Kerr non-linearity using rubidium vapor. AOM: Acousto-optical modulators; GLP: Glan-Laser-Polarizer; NBPS: Non-polarizing beam splitter; HD: Homodyne detector; PZT: Piezoelectric device; M: Mirror. Probe and local oscillator: red beam paths; Control: yellow beam path; Signal: blue beam path. (c) Pulse scheme for determining the relative phase shift experienced by the probe pulses (red solid line) under EIT slowdown (yellow solid line) and N-type conditions (green solid line).}
\end{figure*}

The success of such deterministic, multi-field gates is contingent on two stringent conditions: the first is achieving a large cross-talk between quantum-level fields~\cite{Chang2014}.  The second is the generation of relative non-linear phase operations that can act on either discrete qubit variables or continuous quadrature modes \cite{Braunstein2003}. Different avenues are currently being pursued to overcome the first obstacle. One is the use of highly excited Rydberg states, that have demonstrated the promise for creating a sizeable non-linear medium for individual photons to propagate and interact \cite{Peyronel2012,Firstenberg2013}. Another is the use of strong light-matter interaction, as provided by cavity quantum electrodynamics (QED)\cite{Reiserer2014_2,Vuletic2011,Vuletic2013,Tiecke2014,Scheucher2014}.

Addressing the second condition remains a challenge, yet is key to building truly functional quantum logic gates. Previously, electromagnetically-induced transparency (EIT) enhanced Kerr non-linearities utilizing atomic four-level schemes \cite{Imamoglu1997,Zhu2003} have been used to realize small phase shifts for few photon level classical fields using atomic ensembles \cite{Lukin_gate2009,Chen2011,YuLo2011,Hosseini_2012}.  Although fundamental questions concerning how the non-instantaneous behavior of such Kerr non-linearities preclude the creation of overall cross-phase modulation phase-shifts still persist \cite{Shapiro2006,Banacloche2010}.

Therefore, the availability of evaluation tools that allow the direct characterization of these deterministic phase-shift operations so one can know \emph{a priori} its function regarding specific quantum optical states inputs would be indispensable. Accomplishing such a task in a tractable manner could be attained with relatively simple technological methods, namely EIT non-linearities in room temperature atomic vapor \cite{Pack2006,Pack2007,Li2013,Zhu2014,Li2014} in combination with full quantum optical characterization via coherent state quantum process tomography (csQPT) \cite{Lobino2008,Anis2012}.

Here we report the first complete quantum optical characterization of an engineered phase shift triggered via atomic Kerr-non-linearities. We use a room temperature setup capable of optically controlled phase shifts of $\sim\pi/2$ radians acting on single-photon level probe pulses, triggered by a classical signal field. Our system is realized in a vapor of $^{87}$Rb atoms under the conditions of EIT in a N-type energy level scheme coupled by three optical fields.  We probe our system with weak coherent states and measure the phase and amplitude quadratures of the input and output via time domain homodyne tomography allowing us to perform quantum state reconstruction and directly compare the corresponding density matrices.  Further, by collecting the output data for a sufficient set of weak coherent state inputs, we gain the information needed to completely characterize the phase shift process by csQPT. The resulting process reconstruction yields a rank-4 process super-operator in the Fock states basis which can then be utilized to find out how our phase shift process will behave on arbitrary quantum optical states, either in the discrete or continuous variable regimes.

\textit{Experimental Setup:} We start with a three-level, Lambda atomic EIT scheme composed of two hyperfine grounds states that couple to a common excited state by a weak probe field and a strong control field. The N-type scheme is completed by a third, signal field that couples one of the original Lambda ground states to a separate excited state (see Fig. 1a).  This signal field induces a cross-phase modulation on the probe giving rise to a relative optical phase shift between the fields  \cite{Imamoglu1997,Zhu2003}.

For the primary Lambda system, we utilize two external cavity diode lasers phase-locked at 6.8 GHz to correspond to the $^{87}$Rb ground state splitting.  The probe field is situated 180 MHz red detuned from the $\ket{5S_{1/2}, F = 1 }\leftrightarrow\ket{5P_{1/2}, F = 1 }$ transition with the control field set to the $\ket{5S_{1/2}, F = 2 }\leftrightarrow\ket{5P_{1/2}, F = 1 }$ transition; both lasers are at a wavelength of 795 nm. Lastly, for the signal field we have an additional, third diode laser set to the $\ket{5S_{1/2}, F = 2 }\leftrightarrow\ket{5P_{3/2}, F = 3 }$ transition at 780 nm (Fig 1a).  In our measurements, the signal field is red-detuned by 11.9 MHz.  Furthermore, the control and signal field are fixed to have the same linear polarization with the probe field set to the orthogonal polarization to allow for convenient separation of the fields after atomic interaction.  All fields are also spatially mode matched to a waist of $\omega_0\approx200\mu m$ in a single-rail configuration (see Fig. 1b).  In the pulsed regime the intensity of the probe and control fields are temporally modulated by means of acousto-optical modulators (AOM).  For our medium we use a 7.5 cm long glass cell with anti-reflection coated windows containing isotopically pure $^{87}$Rb with 10 Torr of Ne buffer gas kept at a temperature of 336 K. The cell is pumped with a 1.75 mW control field for 25 $\mu$s before a probe pulse of 1 $\mu$s temporal duration is coupled into the vapor.  After 30 $\mu$s, the control field is shut off and a second probe pulse is sent through the vapor cell to serve as a reference (see Inset of Fig. 1c).  The repetition rate of the entire experiment was 25 kHz.

\begin{figure}[htb]
\centerline{\includegraphics[width=1.0\columnwidth]{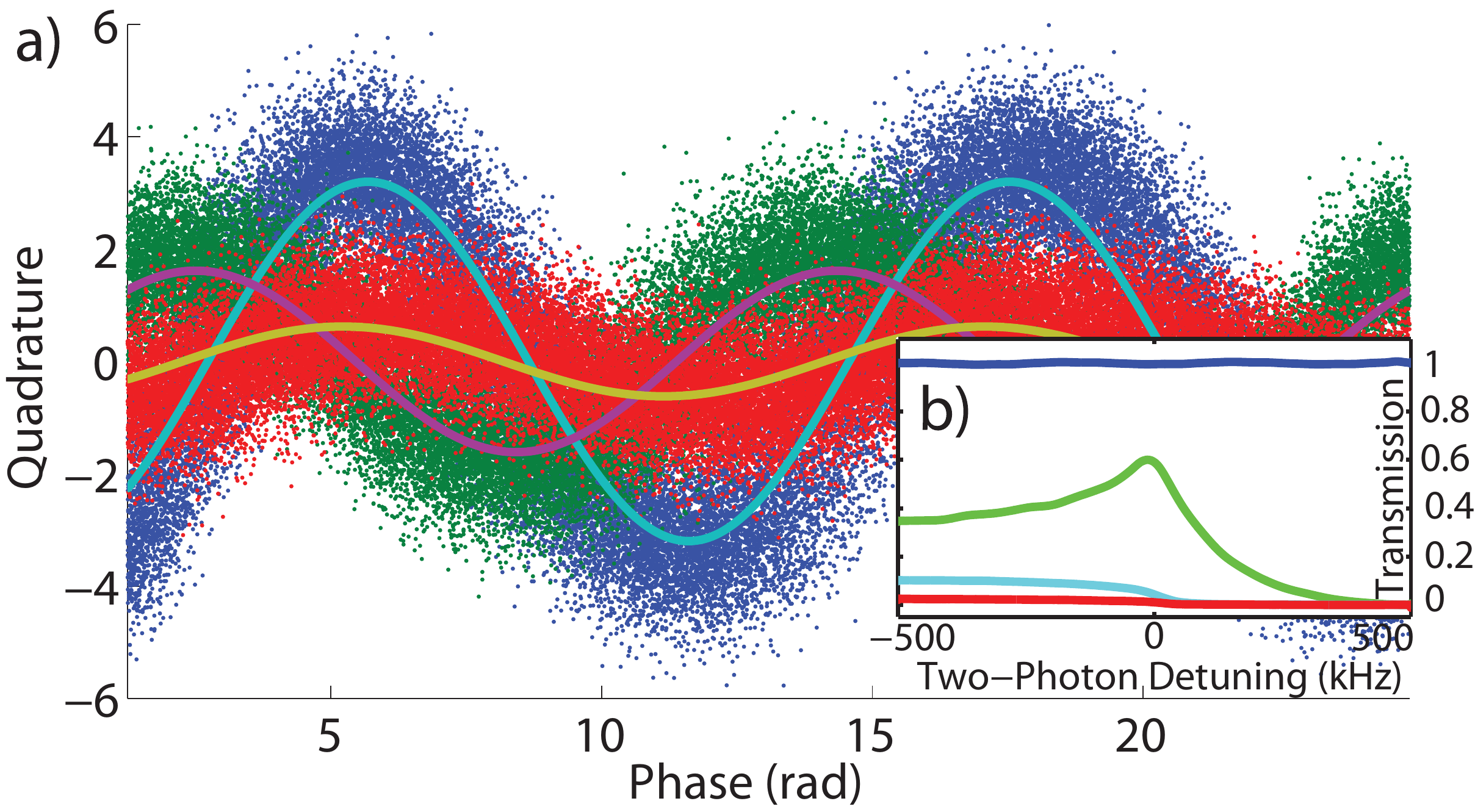}}
\caption{\textbf{Homodyne tomography of Kerr-induced optical phase shift.} (a) Input coherent state (blue dots), phase shifted state under EIT conditions (green dots), state under N-type conditions (red dots) as measured by the homodyne detector, together with their respective fittings for the phase information (solid lines). Note that each dot represents the time integration of a single pulse. (b) Frequency response of the system for the input (blue line), EIT conditions (green line) and N-type configuration with 0.5 mW signal power (light blue line) and 2.1 mW signal power (red line). }
\end{figure}

When only the control field is present, the probe pulse sees a characteristic EIT transparency (see Fig. 2b, solid green line) causing it to experience both slowdown conditions and an optical phase shift.  The presence of a mode-matched signal field can then negate this EIT effect (see Fig. 2b, solid blue line).  Furthermore, the phase shift undergone by the probe pulse can be controllably reduced by applying a particular power of the continuous wave signal field, as changes in transmission correspond to strong dispersion modifications in the atomic medium.

The magnitude of this phase shift is quantified using time domain homodyne tomography by measuring the electric field distribution at all phases \cite{LvovskyRaymer2009}.  To do so, we interfere the weak probe with a strong local oscillator (LO) field derived from the same laser supplying the source of the probe field.  Prior to interference, the local oscillator is reflected off a mirror attached to a piezoelectric device to permit control of the phase and allow total tomographic reconstruction.

In order to calibrate the temporal positions of the electric field quadratures, we first recorded the intensity of bright classical pulses undergoing the different measurement conditions on a photodetector (see pulses in Fig.~1c). From this data, masks delineating the temporal positions of the probe's electric field envelope were generated.  A weighted integration over the homodyne signal using the electric field envelope for a given pulse then yields its time averaged quadrature value.  For all input pulses, those subjected to EIT slowdown and the N-type conditions, we acquired 50 000 phase and quadrature values in the time domain by means of an oscilloscope.  The optical phase shift experienced by the probe pulses under EIT and N-type conditions can then be directly quantified by fitting this homodyne tomography data.  This data is then entered into a maximum likelihood algorithm in order to generate the quantum state or to perform quantum process reconstruction in the Fock basis (see Fig. 2a).

\begin{figure}[]
\centerline{\includegraphics[width=1.0\columnwidth]{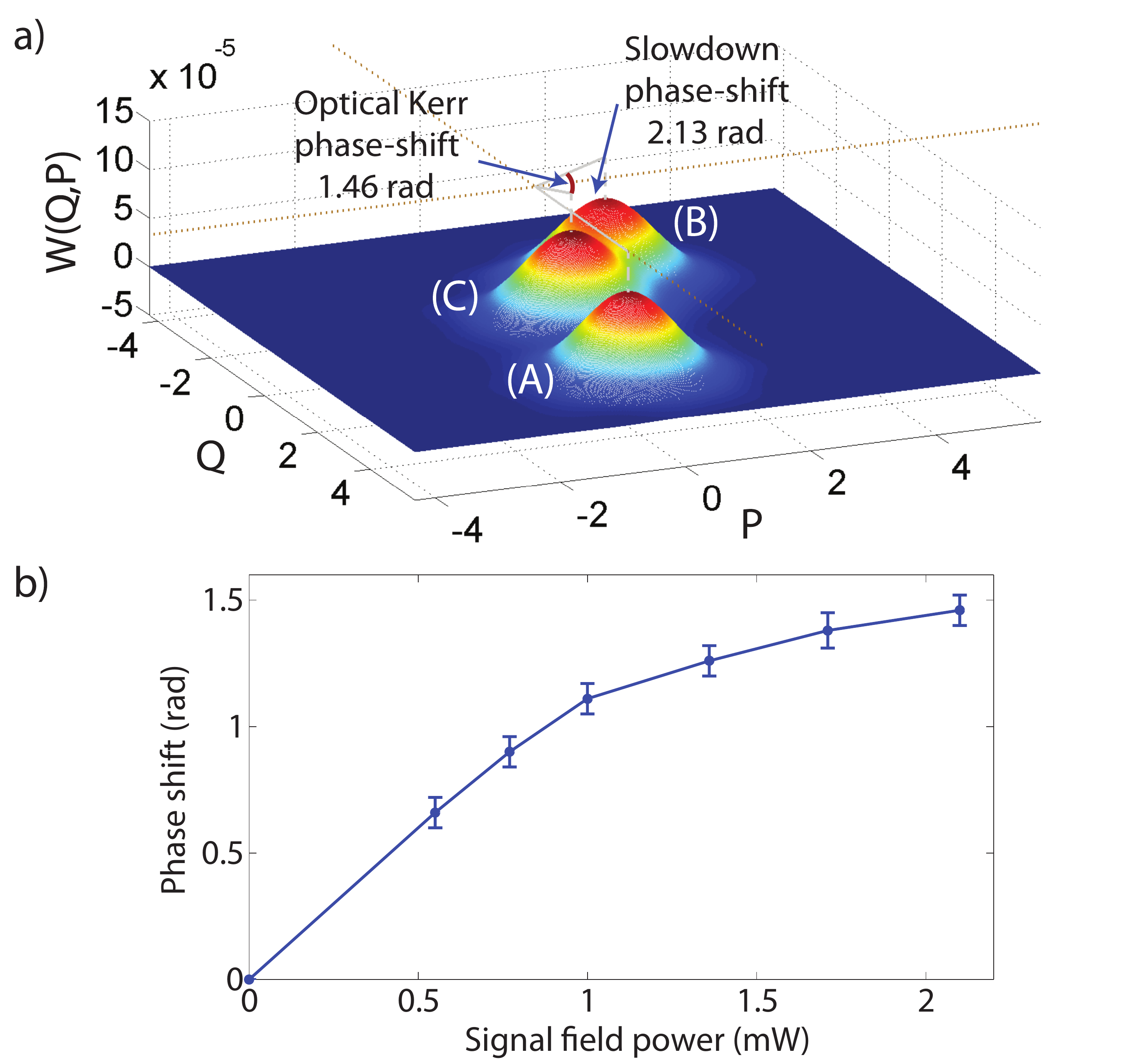}}
\caption{\textbf{Wigner function reconstruction of coherent states under EIT and N-type conditions.} (a) Wigner function reconstruction of the original input state (A), Wigner function reconstruction of the phase shifted state under EIT conditions (B) and Wigner function reconstruction of the N-type modified state using a 2.1 mW signal field (C). (b) Relative phase shift between the EIT and N-type conditions controlled optically as a function of the signal field power. The solid blue line is a guide to the eye. Error bars are statistical.}
\end{figure}

\textit{Discussion:} The phase shift undergone by a single state when subjected to EIT conditions can be understood through direct visualization of the state's Wigner function in phase space.  This can be found by first entering the phase and quadrature data into a maximum-likelihood algorithm to obtain the state's density matrix \cite{Lvovsky2004} followed by direct calculation of the Wigner function~\cite{Leonhardt}.  The information granted by these types of measurements can extend on previous studies \cite{Zhu2003,Chen2011,YuLo2011} as we now attain the full density matrices of the output, which can be compared directly to that of the input. As an example, for an input coherent state containing a mean photon number of $\langle n\rangle=5.4$ ((A) in Fig. 3a), we measure a phase shift of $\Delta\theta_{EIT} = 2.13 \pm 0.04$ rads when subjected to EIT slowdown ((B) in Fig. 3a).  Moreover, we measure the decrease of this phase shift in the N-type scheme when a 2.1 mW signal field is present ((C) in Fig.~3a).  In this case, the phase shift with respect to the input is found to be  $\Delta\theta_{N\mhyphen type} = 0.67 \pm 0.04$ rads.  Hence, the optically induced phase difference between the two schemes is $\Delta\theta_{N\mhyphen type} -\Delta\theta_{EIT}=1.46 \pm 0.06$ rads. From this analysis, we can also find the mean of all quadrature variances $\bar{\sigma^2}$ of the input and output states.  In this case, the variance values corresponding to the input, slowdown and N-type schemes were found to be 0.517, 0.562, 0.517 respectively.  We can see a small thermalization of the slowdown state which is then reduced in N-type conditions as a result of the losses.  Note that due to the losses experienced by the probe in our EIT schemes, the phase shifted Wigner functions are now situated closer to the phase space origin than the original input state.  Here, the transmissions of the EIT and N-type schemes averaged 25\% and 3.5\% respectively.

\begin{figure*}[]
\centerline{\includegraphics[width=2.0\columnwidth]{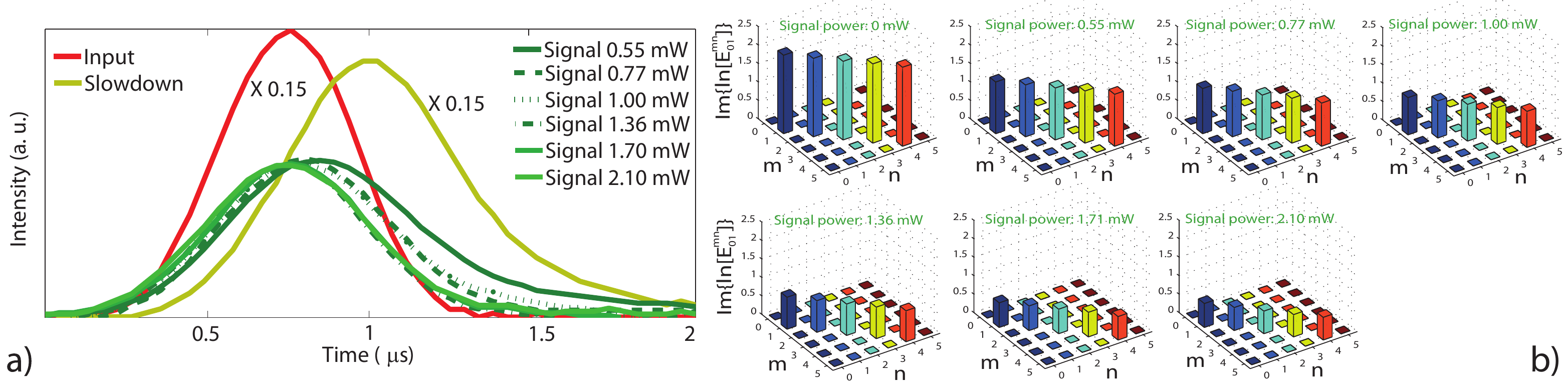}}
\caption{\textbf{Phase transforming elements of the reconstructed process superoperator.} (a) Slowdown pulses with the group velocity modified by different powers of the signal field and represent the temporal mode of the reconstruction. The signal field power ranges from 0.55 mW to 2.10 mW. (b) Plotted are the $Im\{\ln[\mathcal{E}_{01}^{mn}]\}$ elements for different powers of the signal field, where $m$ and $n$ represent the input indices in the Fock basis.}
\end{figure*}

We now turn our attention to characterization of the entire Kerr-based optical phase shift process through means of coherent state quantum process tomography (csQPT) \cite{Lobino2008}.  This is a procedure akin to a ``blackbox'' problem but in the quantum domain, i.e. given any arbitrary but known quantum state $\hat{\rho}$ we can predict $\hat{\mathcal{E}}(\hat{\rho_{i}})$, where $\hat{\mathcal{E}}$ represents our process.  Quantum process tomography is derived from the fact that all quantum processes are linear in the space of density matrices in the Hilbert space defining the process.  More specifically, the technique of csQPT uses coherent states from a common laser source to produce a set of ``probe'' states spanning the Hilbert space where our process will be characterized.  Therefore, by subjecting an ample number of  coherent states $\ket{\alpha_i}\bra{\alpha_i}$ to our optically controlled phase shift and recording the corresponding output  $\hat{\mathcal{E}}(\ket{\alpha_i}\bra{\alpha_i})$, we are provided with the needed information to completely characterize our process in a relatively simple and robust manner.

To accomplish csQPT experimentally, we measured the output states of both the EIT slowdown and N-type processes along with their corresponding inputs for 13 different coherent states with input amplitudes $\alpha_i$ ranging from 0 to 3.3.  This was done for 6 different signal field powers each serving as their own separate characterization. Hence, the temporal modes for every set of conditions was independently measured to account for the group velocity and loss experienced by the probe at a particular signal field power.  Note that the parameters used to achieve our base EIT slowdown conditions remained constant at all times.

The data obtained through homodyne detection was binned into 40 phase and 40 quadrature bins to encompass the entire range of the coherent state measurements.  The binned data was then entered into a maximum-likelihood reconstruction algorithm in line with the procedures described in \cite{Anis2012} to obtain the process super-operator.  Our process was phase invariant, meaning that any fixed phase difference between two input states was preserved after the process and allowed a majority of the elements to be extinguished. We found 100 iterations to be adequate for our reconstruction to converge followed by a truncation of our process superoperator at a maximum photon number at $n=6$.  Note that the linear losses not associated with the process (detector efficiencies, visibilities, etc.) experienced by the probe states were quantified and corrected for in the same manner as was done in prior studies \cite{Kumar2013}.

Our process reconstruction provides a mapping of the input density matrix elements to the resulting output elements holding both the phase and photon number information. While the on-diagonal elements $\mathcal{E}_{kk}^{mm}$ exhibiting the photon number information have been shown in prior studies \cite{Lobino2008,Lobino2009,Kumar2013}, the off-diagonal elements holding the phase information have received less attention \cite{Lvovsky2014} but are of higher relevance in this study. This information can be extracted from the components relating how the input density matrix space maps to a particular off-diagonal element of the output.  The pertinent information can be visualized by taking specific slices of the process tensor, where for instance, the phase value for the $\rho^{out}_{01}$ output density matrix element is related to the $ Im\{\ln[\mathcal{E}_{01}^{mn}]\}$ process tensor elements.  Figure 4b shows these tensor elements for different powers of the signal field where the height of the elements indicates the size of the phase shift relative to the original input state.  In general, it is the summation over the product of these elements with their density matrix counterparts that yields the phase of the output element given by

\begin{equation}
\phi_{kl}=Im\left( \ln\left(\sum_{m,n} \mathcal{E}_{kl}^{mn}\rho_{mn}^{in}\right)\right).
\end{equation}
As expected, the phase component is independent of the input amplitude $\alpha$ (see Fig. 4) over the subspace in which the process was characterized.
The errors in the process tensor elements shown in Figure 4b are those of our experimental statistics and were found to be about $\pm0.06$~rads in all measurements. Other errors include the statistical nature of our quadrature measurements and how they pertain to the quality of our process tensor reconstructions. To evaluate these effects, we simulated multiple data sets by random variation of the quadrature counts at a given phase within its standard deviation.  We then reconstructed set of simulated process tensors $\mathcal{E}_{sim,i}$ and calculated their fidelity $F(\mathcal{E}_{sim,i},\mathcal{E})$ with respect to to our original reconstruction $\mathcal{E}$ using the Jamiolkowski state representation \cite{Jamiolkowski1972}.  We found these fidelities to be near unity with none of the tensor elements shown in Figure 4b deviating more than 0.5\% of the original value.

\begin{figure*}[htb]
\centerline{\includegraphics[width=1.6\columnwidth]{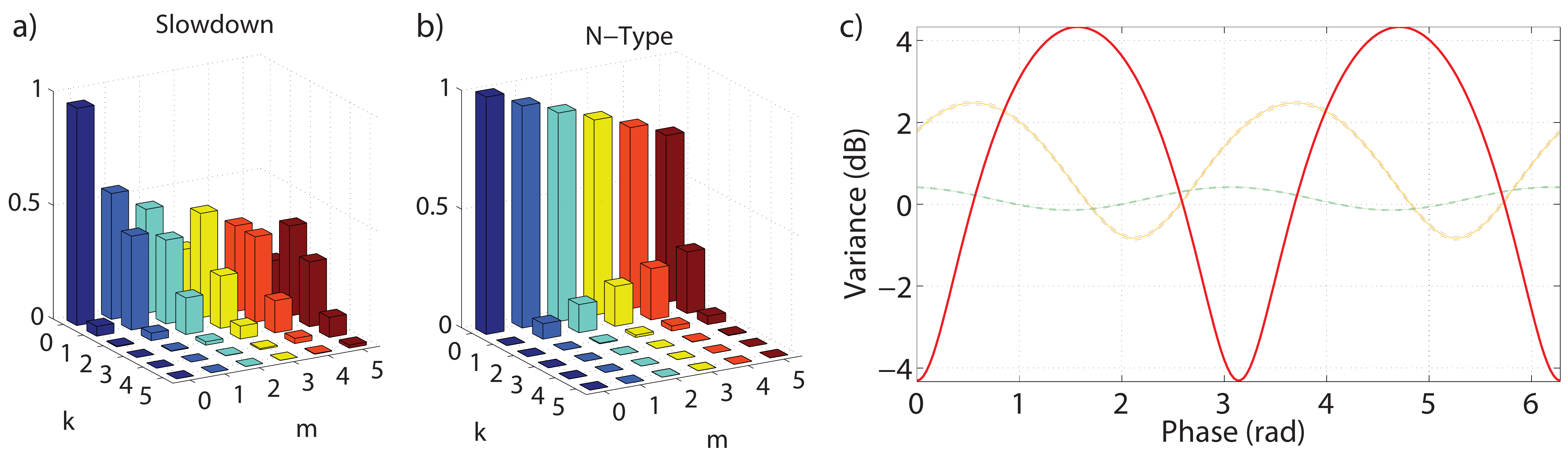}}
\caption{\textbf{Quantum process tensor predictions applied to squeezed light.} Full information about action of the process given by the on-diagonal tensor elements $\mathcal{E}_{kk}^{mm}$ which show the attenuation for the (a) Slowdown and (b) N-type cases.  (c)~Phase shift and quadrature noise information highlighted by the variance versus phase for the theoretical input (solid red line) and the predicted output for slowdown (yellow line) and N-type (green line) scheme with errors (dotted-lines).}
\end{figure*}

The power of csQPT is the ability to make predictions about the process output, or specifically in our case, know \emph{a priori} how our optical phase shift and losses modify input quantum optical states. To exemplify these capabilities, we acted the process tensor reconstructions corresponding to the weakest signal field power of 0.55~mW on a on a quantum optical state in the form of a theoretical squeezed vacuum with $+/-$~4.3~dB of anti-squeezing and squeezing. For the EIT slowdown we found the predicted output state exhibited $\sigma_-^2=-0.83\pm0.04$~dB and $\sigma_+^2=2.47\pm0.04$~dB in the squeezed and anti-squeezed quadratures respectively with a phase shift of $\Delta\theta_{EIT} = 2.12 \pm 0.03$~rads with respect to the input.  Likewise for the N-type scheme, our prediction yielded values of $\sigma_-^2=-0.15\pm0.06$~dB, $\sigma_+^2=0.43\pm0.06$~dB and a phase shift of $\Delta\theta_{N\mhyphen type} = 1.48 \pm 0.03$~rads.

Finally, note that our experimental gate characterization can be applied to any quantum phase gate architecture including those involving discrete variables.  For instance, a characterized phase rotation of $\phi$ means that our process superoperator would act on a Fock state qubit such that $(|0\rangle+|1\rangle)/\sqrt{2}\rightarrow(|0\rangle+e^{i\phi}|1\rangle)/\sqrt{2}$, and hence, csQPT could be an effective tool to determine how qubit variables are modified, thus the bench marking of quantum gate operations becomes a possibility.

In summary, we have characterized a Kerr non-linearity process in the form of an optically controlled phase shift operating at room temperature via the method of csQPT. Our study signifies the first time that such an optical gate-like operation, key for creating quantum optical information processors, has been fully characterized quantum mechanically.

The csQPT technique is achieved with relatively simple optical measurements by probing our phase shift system with only a sufficient set of weak laser pulses and measuring the corresponding output.  Further, we highlighted the potential for this method by using our process reconstruction to predict the effect of our phase shift on a squeezed vacuum state. This represents the first time csQPT has been applied to an optical phase shift-type process and likewise, multi-field interactions.  The simplicity and robustness or this characterization procedure would make it ideal for the facilitation of practical quantum optical gates into future networks and provide an universal tool for the characterization of multi-state quantum components.

As an outlook, we envision bringing our warm vapor implementation towards a true photon-photon phase gate using higher optical densities in a confined volume, as obtained for example by using cavity electromagnetically induced transparency in room temperature atoms \cite{Wu2008}. Quantum gate operation could also be envisioned by combining cavity and dual-rail polarization qubit operation~\cite{Kupchak2015}.

This work was supported by the US-Navy Office of Naval Research, grant number N00141410801 (instrumentation) and the National Science Foundation, grant number PHY-1404398 (personnel and materials). The authors kindly thank A. I. Lvovsky for sharing his maximum likelihood QPT algorithm. C. K. acknowledges financial support from the Natural Sciences and Engineering Research Council of Canada. Correspondence and requests for materials should be addressed to E.F. (eden.figueroa@stonybrook.edu).

\end{document}